# Prediction of under pickling defects on steel strip surface

*Valentina Colla, Nicola Matarese and Gianluca Nastasi

PERCRO, Istituto TeCIP - Scuola Superiore Sant'Anna, Pisa, Italy

*colla{n.matarese, nastasi}@sssup.it*

**Abstract.** An extremely important part of the finishing line is the pickling process, in which oxides formed during the hot rolling stage are removed from the surface of the steel sheets. The efficiency of the pickling process is mainly dependent on the nature of the oxide present at the surface of the steel, but, also, on process parameters such as bath composition and time duration are relevant. When acid concentration, solution temperatures and line speed are not properly balanced, in fact, sheet defects like under pickling or over pickling may happen and their occurrence does have a very serious effect on cold-reduction performance and surface appearance of the finished product. Furthermore, product damage from handling or improper equipment adjustment can render the steel unsuitable for further processing. This is the reason why it is important that process significant parameters are controlled and maintained as accurately as possible in order to avoid these undesired phenomena. In the present work, a control algorithm, composed by two different modules, i.e. decision tree and rectangular Basis Function Network, has been implemented to aim of predicting pickling defects and suggesting the optimal speed or the admissible speed range of the steel strip in the process line. In this way the most suitable line speed value can be set in an automatic way or by the technical personnel.

## 1. Introduction

To attain a complete elimination of the oxide from steel strips surface a mandatory production step - the pickling process - is needed in order to make further treatments (e.g. hot deep galvanizing) and to maintain quality standards. One of the main problems during this process is the difficult evaluation of the pickling state. If on the one hand it is evident that high strip quality requires a complete elimination of the oxide on the strip surface, thus under-pickling cannot be tolerated, on the other hand a too long permanence of the strip in the pickling bath can lead to an erosion of the steel strip surface itself (over-pickling). Both these problems lead to more expenses or losses: in facts under-pickled strips need to be re-processed, while the quality of over-pickled ones must be downgraded. The main issue in the detection of over and under pickling defects by means of historical data analysis lies in the unbalance of correctly pickled strips, which are the most, and the class of defective ones, which makes the application of standard classification methods ineffective.

In details, after hot rolling at the steelworks, the steel strip passes without interruption through a cooling process, where the steel strip is cooled to the desired temperature by spraying water on both sides of the steel. After water cooling, the strip temperature is decreased at a slower rate by cooling in air. Oxide scale is formed on the surface of the steel during the entire cooling phase.

The oxide scale has to be removed from the steel surface before further processing of the steel strip, for instance cold rolling. Furthermore, the influence of defects and their removal, must be considered when manufacturing to specifications that relate to certain surface quality requirements. Removal of oxide is carried out within mechanical methods [1,2,3] including polishing method and shot blast method, particularly used for removing dense and adherent oxides scales from stainless steel [4,5]; and chemical methods like electrochemical pickling [6], or chemical pickling methods.

This work is focused on steel chemical pickling bath method [7,8] (namely pickling), which is the part of the finishing process in the production of steel products in which oxide and scale are chemically removed from the surface of strip steel in order to obtain a clean surface. During the pickling process, steel strip is pulled through three different tanks that contain baths of aqueous acid solution with different concentration: 5, 10 and 15 wt% of hydrochloric acid (HCl) and subsequently steel strip passes through water rinsing tank, with a cascade flow of pickling solution counter current to the direction of the strip movement. Fig.1 shows the pickling process and the involved variables.

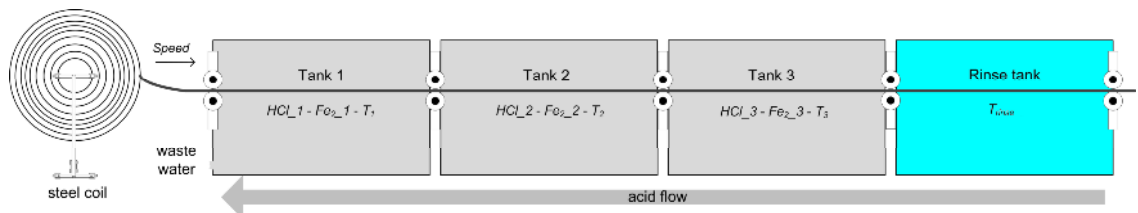

**Figure 1.** Pickling process

In order to prevent and reduce to minimum the iron dissolution from the steel strip, allowing a preferential attack on the scale, in all the baths organic compounds, namely corrosion inhibitors, are added. The goal of the inhibitors addition is to protect the strip from the attack by the hydrochloric acid solution, by means of protective film that sticks on the surface of the steel strip but not on the oxide. It is fundamental to set the optimum concentration of the corrosion inhibitor, because a suitable trade-off must be reached between its concentration, cost and the need to limit the amount of these hazardous components into the wastewater coming from this process, in order to minimise the process impact on the surrounding environment.

In the literature many methods has been studied in order to evaluate the concentration of corrosion inhibitor, for setting its optimal level while keeping its action effective and minimizing costs and dissolution of hazardous material into wastewater [9].

The objective of this research is to identify the conditions when the under pickling phenomena take place, in order to firstly detect and predict their formation. In this paper a model composed by two different modules, a decision tree and a Rectangular Basis Function Network (recBFN), is described. Its aim is to suggest the optimal speed or admissible speed range for the process line, given steel strip parameters (e.g. strip width and thickness, steel grade) and acid bath conditions, with the objective to reduce at minimum the occurrence of under pickling defects. The implemented overall model achieved noticeable results.

The paper is organized as follows: in Sec. 2 the definition of the problem and parameters are depicted; Sec. 3 describes the implementation of the proposed model; Sec. 4 depicts the obtained results and finally Sec. 5 provides some concluding remarks and future work.

## 2. Problem definition

The pickling process, if not suitably performed, can lead to the formation of defects on the surface of the steel strip, which are related to two opposite conditions, often referred as over pickling and under pickling. Over pickling results from the line delays which permit sections of the steel to remain in the acid too long. The presence of an inhibitor reduces iron loss, but when an inhibitor is not used, iron loss during a short delay period appreciably reduces thickness of the steel and raises the hazard of hydrogen embrittlement. This kind of defects must be prevented by adding inhibitor to the acid solution and also by increasing the line speed [10], if possible. On the other hand, under pickling occurs when the acid solution does not completely dissolve the oxide, due to a low concentration of the acid solution and/or an excessive line speed, which shorten the time of permanence of each portion of the strip in the acid solution. The present work is particularly focused on the avoidance of the under pickling phenomenon.

The goodness of pickling process is affected by many variables such as the steel composition, acid solution, scale structure, concentration of corrosion inhibitor, baths temperature and line speed, as studied in a previous work [11]. Usually, under pickling is the direct results of a lack of control over the above-cited process variables, thus many studies have been proposed on control of pickling process that use neural network model [13,14] or a hybrid grey box modelling [15], in order to model the steel pickling process. The use of neural networks and grey box models is appropriate as there exist some, although incomplete, knowledge about the considered process that can be exploited. In the present work, the aim of the proposed artificial intelligence-based approach is not only to predict the under pickling occurrence but also to provide indications for avoiding such phenomenon.

A dataset deriving from a theoretical model of the pickling process, which was developed by ArcelorMittal Research [16,17], have been exploited. The support of the industrial technical personnel has been important both for setting the model parameters and for selecting the variables that mainly affects the occurrence of the under pickling phenomena.

The selected variables and their nomenclature are listed below with a subdivision in physical and process parameters:

- Physical coil parameters
    - $W$ – coil weight
    - $t_s$ – strip thickness
    - $w_s$ – strip width
- Pickling process parameters
    - $T_1$ – temperature of tank 1
    - $T_2$ – temperature of tank 2
    - $T_3$ – temperature of tank 3
    - $T_{rinse}$ – temperature of rinse
    - $v$ – speed of pickling line
    - $HCl\_1$ – acid concentration of tank 1
    - $HCl\_2$ – acid concentration of tank 2
    - $HCl\_3$ – acid concentration of tank 3
    - $Fe_2\_1$ - $Fe_2$ concentration of tank 1
    - $Fe_2\_2$ - $Fe_2$ concentration of tank 2
    - $Fe_2\_3$ - $Fe_2$ concentration of tank 3
    - $Under\_P$ – under pickling defects

The initial dataset, composed by about 2000 records, has been pre processed by purging outliers and non valid entries, obtaining a final database composed by about 1800 records. The percentage of data referring to the occurrence of under pickling defects and to the absence of pickling defects is of 75% and 25%, respectively.

The noticeable difference between the number of data belonging to the two classes makes the dataset quite unbalanced and prevents the application of traditional classification methods as well as standard soft-computing based classification approaches.

In order to minimize the under pickling phenomena on the surface of the steel strip, the fastest way to govern the effectiveness of the considered process is to set a right value of the speed of steel strip, which affects the permanence time of the steel strip in the pickling bath and the effect of both the hydrochloric acid solution and the corrosion inhibitor on the steel strip surface [10].

## 3. Model implementation

The model proposed in this work is composed by the union of two different modules, a decision tree-based model [18] and a RecBFN-based model [19]. The variables *underP* and *v* are the targets (i.e. they must be predicted), while some of the others are fed as input in the implementation of the developed model.

The first module aims at defining a valid process window, i.e. a range of process parameters ensuring a good behaviour of the pickling line. A decision tree model is used, by classifying the input parameters with respect to the associated value of the line speed, by considering only the data contained in the available dataset, that are related to products not affected by pickling defects. The final aim of such system is to suggest to the operator a valid range of the line speed that allows to avoid the under pickling and over pickling defects on the basis of some coil and process parameters.

The second module is a particular kind of neural network and its presence is due to the fact that, when the control is applied on the real line and working at full stretch, there is a decreasing number of under pickling defects on the strip steel surface, because the capability of the control system predict and adjust the speed line will improve through time. Considering this declining frequency of the under pickling defects, a classical learning method which is usually applied to balanced datasets (i.e. datasets where the number of the data belonging to the different classes are balanced) could not be able to detect the anomalous conditions which give raise to under pickling. Therefore a model based on soft computing techniques, which is specialized on imbalanced datasets and is composed by a neuro-fuzzy model called RecBFN (rectangular Basis Function Network) [20,21], has been implemented and subsequently integrated in the decision tree module.

The following two subsection are devoted to a more detailed description of the two above-mentioned modules, while the third subsection describes how the two modules have been integrated and how the global model works in practice.

### a. Decision Tree module

A decision tree [18,22] is a type of tree-diagram that is used for determining the optimum course of action, in situations that are characterised by several possible alternatives with uncertain outcomes. The resulting tree displays the structure of a particular decision and the interrelationships and interplay between different alternatives, i.e. decisions, and possible outcomes. They are widely adopted in the field of operations research and decision analysis, in order to point out the strategy or the procedure that are most suitable to reach a predefined objective. In detail, at each node of the tree, the algorithm chooses one attribute of the data that most effectively splits the set of samples into subsets enriched in one class or the other. Its criterion is the normalized information gain that results from choosing an attribute for splitting the data. The attribute with the highest normalized information gain is chosen to make the decision.

In the present case, a classification algorithm based on a decision tree has been applied to the aim of identifying a valid process window i.e. an optimal range for the line speed value which avoids the occurrence of both under pickling and over pickling defects. The following variables have been selected in the implementation phase of this model, which are related to both product and process: $t_s$, $W$, $T_1$, $T_2$, $T_3$, $T_{rinse}$, $HCl\_1$, $Fe_2\_1$, $HCl\_2$, $Fe_2\_2$, $HCl\_3$, $Fe_2\_3$.

By considering that the line speed is the only available degree of freedom for technical operators working on the pickling line, the variable $v$ parameter is assumed as target variable in the classification phase. In order to classify the input variables with respect to the target, all the line speed values, that range from 104 m/s to more than 385 m/s, have been subdivided into the following three different classes of same size:

- class A: $104 < v \leq 244$
- class B: $245 < v \leq 384$
- class C: $v \geq 385$

Moreover a class U of defective coils has been added, in order to take into account those coils where the under pickling phenomenon is previewed to occur independently on the value set for the line speed (for instance because of the low concentrations of hydrochloric acid in the tanks and/or unsuitable temperature values). A binary tree has been built on the defined training set, by means of a decision tree algorithm, where each node represents an input variable and the leaf node represent one of the three previously defined class i.e. the predicted class of the target variable $v$. Following each branch of the constructed tree, the model is capable to identify a valid process window, showing the line speed range that should be set in the considered process condition in order to avoid under pickling and over pickling defects.

### b. RecBFN module

In order to correctly detect and classify defective coils also when a high unbalance level will be reached among defective and non-defective coils due to the implementation of the speed line control at full stretch, a neuro fuzzy model called Rectangular Basis Functions Network [17] has been implemented. This kind of neural network, which is depicted in Fig. 2, consists of hidden units, each covering a rectangular area in the input space, using a trapezoidal activation function.

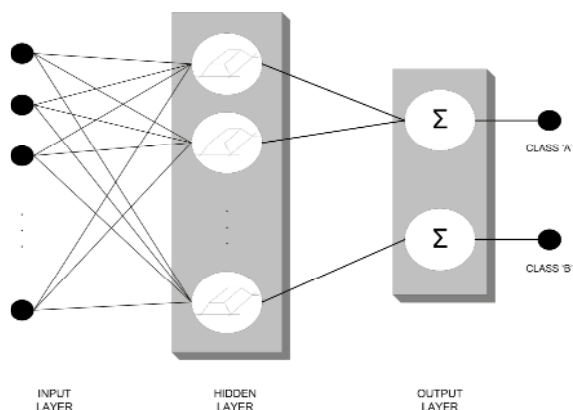

**Figure 2.** recBFN structure

In detail, in the hidden layer of the network, each neuron represents a fuzzy point and it is associated with the two classes, i.e. under pickling defects and no defects. Each neuron takes as input a vector containing the selected parameters computing the degree of membership of the input pattern to the fuzzy point. Lastly the output layer assigns a class to the input pattern processing the input information. The underlying training algorithm allows easy and fast construction of these types of networks and no parameters need to be adjusted, only normalization of the input-data is necessary.

This kind of neural network has been chosen because it works well on unbalanced datasets [20,21]. Among the pickling process variables, the parameters defined in the testing phase, which are used as inputs of the model are: $T_3$, $HCl\_1$, $Fe_2\_1$, $HCl\_2$, $Fe_2\_2$, $HCl\_3$, $Fe_2\_3$. In order to get a prediction of maximum line speed in the case of under pickling defect, $v$ has been used as an additional free input: the whole range of possible line speeds within a predefined range is scanned starting from the lowest bound as far as the inputs configuration is recognized as corresponding to good working conditions. As an under pickling condition is met, a maximum line speed that avoid under pickling is suggested. In this way, by suggesting the first helpful velocity that would avoid under pickling defects, the occurrence of the opposite phenomena (i.e. over pickling), which is due to an excessive time spent in the pickling tank, is prevented.

## 1.1. Integration of the two modules

The two different units composed by RecBFN model and decision tree model have been integrated as depicted in Fig. 3.

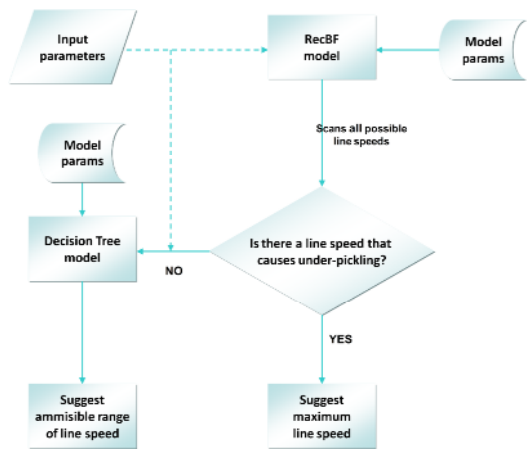

**Figure 3.** Global model flow chart

By using the input variables listed in Par. 3.2, the system firstly runs the RecBFN model in order to evaluate the input parameter against all possible line speed value in a predefined range; if there is a line speed value that, together with other parameters, could cause the under pickling phenomenon, a maximum admissible line speed is suggested. Otherwise the decision tree unit is run, all the input parameters are used to classify the coil with respect to the variable $v$ and a suitable range of line speed values is suggested, in order to prevent the under pickling defects.

## 4. Results

During the implementation phase the dataset has been divided into a training set, which contains the 75% of the total records, and a validation set, that is composed by the remaining 25% of the records. A particular attention has been paid to keep the same proportion between the two existing classes, i.e. defective and non defective coils.

Each of the two modules has been separately designed and trained, as discussed in par. 3.1 and par. 3.2. It must be underlined that the decision tree must be trained only with data referring to standard conditions, while the training of the neural network-based model needs data associated to the situations of no-pickling and under pickling.

Finally a global validation of the model has been executed on the remaining 25% of the dataset, which has not been used in the training phase of both modules that compose the model.

The decision tree-based module has been trained through a standard C4.5 algorithm [22], obtaining a tree of 45 levels, with a size of 89. The test of such tree, that is performed on the validation dataset provided a rate of correct classifications of 75.7%, which means that the decision tree successfully classifies all the inputs with respect to the variable v.

For the RecRBF-based module, very low values of the misclassification rate has been achieved on both training and validation sets, namely 0.23% and 0.30% respectively.

In order to evaluate the performance of the global model the following classical and widely adopted indexes in the classification tasks [23] have been exploited.

**Table 1.** Confusion matrix

|  | Condition | |
|---|---|---|
|  | TRUE | FALSE |
| Test Outcome — POSITIVE | True Positive | False Positive |
| Test Outcome — NEGATIVE | False Negative | True Negative |

Considering the confusion matrix, in Table 1, where the columns represent the correctly classified instances, while the rows represent the real instances of the particular class, the performance indexes are defined as follows:

- **Precision** – TP / TP + FP
  This is a measure of the proportion of the classified matches that are true matches;

- **Recall** – TP / TP + FN
  This is a measure of the proportion of actual matches that have been correctly classified;

- **F-Measure** – 2*Precision*Recall / Precision + Recall
  This is a measure of the compromise between precision and recall;

Table 2 shows the obtained results for the four classes, i.e. classes A, B and C, that group the decision tree classification, and class U that refers to the RecBFN output for the defective coils.

**Table 2.** Global model performance

|  | Precision | Recall | F-Measure |
|---|---|---|---|
| class U: under – pickled | 0.961 | 0.958 | 0.959 |
| class A: (104 - 244] | 0.934 | 0.948 | 0.941 |
| class B: (245 - 384] | 0.913 | 0.961 | 0.936 |
| class C: [>= 385] | 0.938 | 0.682 | 0.789 |

The performance of the global model is noticeably good, because all the values of the adopted indexes are close to one. Thus by means of this model it is actually possible to predict, with a high degree of precision of about 96%, the case in which probably at the end of the pickling process some under pickling defects are formed on the coil surface, by allowing the operator to set a optimal value for v according to the suggestion provided by the model. This is possible thanks to the successful classification of coil and process parameters. The results are computed by taking into account that, when under pickling actually occurs, the fact that predicted line speed is lower than the real one demonstrates the good behaviour of the model, as a decrease of the speed line value can lead to an optimal working by avoiding the under pickling phenomenon.

In addition, the very low rate of false alarms (i.e. coils without pickling that are classified as defective), which is equal to 0.034%, shows that the model has a very high probability to predict the under pickled defects when only a few working presents this undesirable phenomena, by also suggesting a maximum value of the line speed, that allows to avoid the occurrence of under pickling defects without conditioning the over pickling rate, because the suggested velocity value is slightly lower than the real value.

A training phase should be periodically performed on new data, in order to guarantee the global reliability of the overall system over the time.

## 5. Conclusions

During the pickling process, the under pickling and over pickling defects are usually the direct results of lack of control over process variables including acid concentrations, solution temperature and contact time deriving from pickling line speed. Up to date most pickling lines are run by technical operators, which view the surface of the steel strip time to time and adjust process parameters, by mainly decreasing the line speed; this operation does not always achieves the desired results.

The proposed model, composed by two different units, is capable to predict the under pickling occurrence in a pickling process, from coil and process parameters, by also suggesting a defined value for v or its admissible range. By means of decision tree unit, a binary tree has been built allowing a classification of valid process window, from which it is possible to suggest the speed range that should be set in the considered process conditions in order to avoid under pickling defects.

In addition, a neuro-fuzzy model called RecBFN has been used in order to predict the under pickling defects starting from unbalanced dataset, a situation that is likely to occur when the implemented control works at full stretch and a non optimal working decreases due to the self-learning capability of decision tree.

By means of the union of the two different units and starting from coil and process parameters, the global model is able to predict the maximum line speed or the speed range that allows to avoid the under pickling defects.

The only degree of freedom at disposal of the operator during the pickling process is the line speed, which determines the time spent in the tanks by the steel strip and, consequently, how long the acid solution attacks the surface of the steel, with the possibility of formation of under pickling or over pickling defects. The operator will be helped by the suggestion of the maximum line speed or the admissible speed range, that will lead to good working, without the presence of over and under pickling defects, as it is possible to react, by adjusting the speed line so as to carry out the pickling process under optimal conditions, by avoiding the occurrence of under pickling phenomena and obtaining a quality improvement of the final product with a cost reduction.

This model can also be transferred on different line typologies, by simply executing a new training phase of both modules.

As future work, the adaptation of the same global model structure will be done in order to detect over pickling defects, which more rarely happens with respect to the under pickling defects, not allowing at the moment the application of the techniques proposed in this work. The model needs improvements in order to deal with even more imbalanced classes than those present in the under pickling problem. In such way will be possible the optimize the pickling process, allowing at the same time the reduction of both over and under pickling defects.

## Acknowledgement

The work described in the present paper was developed within the project entitled "*Optimized productivity and quality of pickling by on-line control of the pickled surface*" (Contract No. RFSR-CT-2005-00021) that has received funding from the Research Fund for Coal and Steel of the European Community. The sole responsibility of the issues treated in the present paper lies with the authors; the Commission is not responsible for any use that may be made of the information contained therein.

The authors wish to thank all the colleagues who participated to the above-cited RFCS project, Dr. Antonella Dimatteo and Dr. Teresa A. Branca for the fruitful discussions which provided a valuable contribution to the present analysis.